\documentclass[useAMS,usenatbib]{mn2e}
\usepackage{graphicx, setspace, subfigure, latexsym, amssymb, amsmath, booktabs, wasysym, paralist, longtable}
\usepackage{multirow}
\usepackage{gensymb}
\usepackage{verbatim}
\usepackage{siunitx}
\usepackage{upgreek}

\title[HTRU XI]{The High Time Resolution Universe Survey - XI. Discovery of five recycled pulsars and the optical detectability of survey white dwarf companions}
\author[S. D. Bates et al.]
{S. D. Bates$^{1,2}$, D. Thornton$^{1,3}$, M. Bailes$^{4,5}$, E. Barr$^{5,6}$, C. G. Bassa$^7$, N. D. R. Bhat$^{4,5,8}$,\newauthor M. Burgay$^9$, S. Burke-Spolaor$^{10,11}$, D. J. Champion$^6$, C. M. L. Flynn$^{4,5}$, A. Jameson$^{4}$,\newauthor S. Johnston$^3$, M. J. Keith$^1$, M. Kramer$^{6,1}$, L. Levin$^{12}$, A. Lyne$^{1}$, S. Milia$^{9,13}$, C. Ng$^6$,\newauthor E. Petroff$^{3,4,5}$, A. Possenti$^9$, B. W. Stappers$^1$, W. van Straten$^{4,5}$, C. Tiburzi$^{3,13}$\\
$^{1}$Jodrell Bank Centre for Astrophysics, School of Physics and Astronomy, The University of Manchester, Manchester M13 9PL, UK\\
$^{2}$National Radio Astronomy Observatory, PO Box 2, Green Bank, WV 24944, USA\\
$^{3}$CSIRO Astronomy \& Space Science, Australia Telescope National Facility, P.O. Box 76, Epping, NSW 1710, Australia\\
$^{4}$Centre for Astrophysics and Supercomputing, Swinburne University of Technology, PO Box 218, Hawthorn, VIC 3122, Australia\\
$^{5}$ARC Centre of Excellence for All-Sky Astronomy (CAASTRO), Mail H30, Swinburne University of Technology, PO Box 218,\\ \: Hawthorn, VIC 3122, Australia\\ 
$^{6}$MPI fuer Radioastronomie, Auf dem Huegel 69, D-53121 Bonn, Germany\\
$^{7}$ASTRON, The Netherlands Institute for Radio Astronomy, Postbus 2, 7990 AA, Dwingeloo, The Netherlands\\
$^{8}$Iternational Centre for Radio Astronomy Research, Curtin University, Bentley, WA 6102, Australia\\
$^{9}$INAF - Osservatorio Astronomico di Cagliari, via della Scienza 5, 09047 Selargius, Italy\\
$^{10}$NASA Jet Propulsion Laboratory, M/S 138-307, Pasadena CA 91106, USA\\
$^{11}$California Institute of Technology, 1200 E California Blvd., Pasadena, CA, 91125 USA\\
$^{12}$Department of Physics and Astronomy, West Virginia University, Morgantown, WV, 26506 USA\\
$^{13}$Dipartimento di Fisica, Universit\`{a} degli Studi di Cagliari, Cittadella Universitaria, 09042 Monserrato (CA), Italy}
\begin{document}

\date{\today}

\pagerange{\pageref{firstpage}--\pageref{lastpage}} \pubyear{2012}

\maketitle

\label{firstpage}

\begin{abstract}
    
We present the discovery of a further five recycled pulsar systems in the 
mid-Galactic latitude portion of the High Time Resolution Universe (HTRU) Survey. 
The pulsars have rotational periods ranging
from 2\,ms to 66\,ms, and four are in binary systems with orbital periods 
between 10.8\,hours and 9.0\,days. Three of these binary systems are particularly interesting;
PSR~J1227$-$6208 has a pulse period of 34.5\,ms and 
the highest mass function of all pulsars with near-circular orbits.
The circular orbit suggests that the
companion is not another neutron star, so future timing experiments may reveal
one of the heaviest white dwarfs ever found ($>$ 1.3 M$_\odot$).
Timing observations of PSR~J1431$-$4715 indicate that it is eclipsed by its
companion which has a mass indicating it belongs to the redback class
of eclipsing millisecond pulsars.
PSR~J1653$-$2054 has a companion with a minimum mass of only $0.08~\mathrm{M}_\odot$, 
placing it among the class of pulsars with low-mass companions. Unlike the majority 
of such systems, however, no evidence of eclipses is seen at 1.4~GHz.

\end{abstract}


\section{Introduction}
Although pulsars are commonly born with short spin periods \citep[e.g.\,][and references therein]{mel2002},
the spin-down rate is such that pulsars are often observed to have spin periods 
of order $1\mathrm{~s}$. However, the \emph{millisecond pulsars} (MSPs) 
spin with periods $\lesssim 30\mathrm{~ms}$. The MSPs are located 
in a region of the $P$-$\dot{P}$ diagram distinct from \emph{ordinary}
 pulsars.  In order to be rotating so rapidly, MSPs are
 thought to have undergone a \emph{spin-up} phase in their evolutionary 
 history. Spin-up involves
 accretion of matter from an orbiting companion onto the neutron star (NS) 
 \citep{acrs82}, which
begins when the companion star ages and expands, in some cases,
to overflow its Roche lobe.
  During the accretion phase, systems are thought to be observed as X-ray binaries, 
  both as high-mass (HMXB) and low-mass X-ray binaries (LMXB) 
  depending upon the companion mass.

The evolutionary link between X-ray binaries and MSPs was strengthened with the
 discovery of SAX J1808.4$-$3658, an X-ray binary exhibiting periodic intensity 
fluctuations with a period of $2.4\,\text{ms}$ \citep{wv98}.  More recently, 
PSR~J1023$+$0038 was argued to have switched from an LMXB to radio pulsar
 phase \citep{akb+10}, and now has undergone a transition back to an
 LMXB \citep{stappers2013}. The radio pulsar J1824$-$2425I was the first to be observed to
 switch to an LMXB phase for a period of around a month, before radio pulsations 
 were once again detected \citep{pap2013}. All three systems
 further strengthen the spin-up model of MSP formation.  

 Some HMXBs may have a donor sufficiently massive to undergo a core-collapse 
 supernova (ccSN), providing a strong kinematic
 kick to the system.  This kick and mass loss from the companion will either disrupt
 the binary totally or impart a significant eccentricity to the orbit \citep{cb05}.
  What remains is therefore either an eccentric Double Neutron Star (DNS) system, 
for example PSR B1913$+$16 \citep{ht75a} and PSR J0737$-$3039 \citep{bdp+03, lbk+04},
 or two separate NSs.  In both cases one NS has undergone a spin-up phase 
 (the recycled pulsar)
 while the other is a young NS (either of which may be observable as a pulsar).  

The LMXBs are systems where the companion has a lower mass than the pulsar.
The companions reach the end of their lives as a red giant with 
expulsion of the outer layers of the star, leaving a degenerate core --- a white dwarf
(WD).  The long-lived period \citep[$\sim 10^{8}\text{ years}$, ][]{tv03} of stable
 mass transfer acts to circularise the orbit after the ccSN of the NS progenitor
 \citep{bv91}, explaining why Galactic field MSP-WD 
binaries usually have a low orbital eccentricity \citep{phi92}.  

The fact that most MSPs are found in binaries is in agreement with the spin-up 
evolutionary scenario, however $\sim 20\%$ of observed MSPs in the Galactic field are 
isolated.
 Besides the formation of isolated MSPs from disruption of the orbit during 
 a second ccSN as the HMXB phase ends,  MSPs may destroy their companions
 through ablation due to the wind from the pulsar \citep{rst89a}.
  Systems undergoing this process are known as \emph{black widow} systems \citep{fst88}, 
consisting of an MSP with an ultra-low-mass companion 
($\text{M}_{\text{c}} \approx 0.02\:\text{M}_{\odot}$).  Radio emission from the pulsar
 is sometimes observed to be eclipsed by ionised material surrounding the companion 
and, if the orbital inclination is favourable, the companion itself (for example 
PSR B1957$+$20, \cite{fst88}).  
The timescale for total ablation of the companion is, however, too
 long to explain the observed number of isolated MSPs \citep{el88}.  So-called 
\emph{redback} binary systems \citep{rob11}, such as PSR J1740$-$5340A in globular cluster
NGC6397, also exhibit orbital eclipses of the pulsar emission \citep{dlm+01}. 
 redback systems have a more massive companion,
 ($\text{M}_{\text{c}}  \sim 10^{-1} \text{M}_\odot$) 
 often seen to be non-degenerate in optical studies \citep{rob11}.
  These systems have Roche lobe filling factors of $0.1$-$0.4$, 
and as such are possibly an interesting intermediate stage between X-ray binaries
 and Galactic field MSP binary systems \citep{bkr+13}.  

Using the 64-m Parkes radio telescope the HTRU 
survey \citep{kjv+10} is providing a comprehensive search of the entire
Southern sky with high time and frequency resolution for pulsars.
It uses the 20\,cm multibeam receiver which was also used by the Parkes 
Multibeam Pulsar Survey (PMPS) \citep{mlc+01} combined with a 
digital backend system with superior spectral and temporal resolution to the analogue
 filterbank system used in the PMPS.

This paper outlines the discovery and subsequent observation of five 
recycled pulsar systems. Section \ref{s:disc_timing} outlines
the observations taken to discover and observe these pulsars. 
In Section \ref{s:results} we present the results
 of our timing programme to date, including details of the
 pulse profiles and, where possible, multi-frequency observations. Finally, 
in Section \ref{s:opt} we present calculations of predicted WD optical 
brightnesses for all published MSP-WD systems discovered in the HTRU survey.  

\begin{table}
	\begin{center}
        \caption{Observing system details for the timing observations made as part of this work; observing bandwidth, BW, number of frequency channels, $n_\mathrm{chan}$, and mean observation length, $\langle\tau_{\text{obs}}\rangle$. Note the specifications for the Lovell telescope take into account the removal, as standard, of a section of the observing bandwidth due to contamination by RFI.}
		\begin{tabular}{lcccc}
		\hline
		Telescope  & Centre Freq. & BW & n$_{\text{chan}}$ & $ \langle\tau_{\text{obs}}\rangle $ \\
		&(GHz) &(GHz) & &(s)\\
		\hline
		Parkes 64-m & 0.732 & 0.064 & 512 & 900 \\
		& 1.369 & 0.256 & 1024 & 600 \\
		& 3.094 & 1.024 & 1024 & 900\\
		\hline
		Lovell 76-m & 1.524 & 0.384 & 768 & 1800 \\
		\hline
		\end{tabular}
		\label{table:timing}
	\end{center}
\end{table}

\begin{table*}
    \caption{Observed and derived parameters for the five MSPs. The DM distance has been estimated using the NE2001 model of \citet{ne2001}, while a pulsar mass of $1.4\,\mathrm{M}_\odot$ has been assumed in calculating companion masses. Estimates of the contribution to the period derivative from the Shklovskii effect have been included, where significant proper motions have been measured. ELL1 refers to the timing model outlined in the appendix of \citet{lange2001}, and ELL1H is a modification to this model to allow the Shapiro delay to be parameterized in terms of $h_3$ and $h_4$ (see text and Equation~\ref{e:shap}). Note all parameter errors have been multiplied by $\sqrt{\chi^2}$.}
		\begin{tabular}{lccccc}
		\toprule
		Parameter               & J1227$-$6208 & J1405$-$4656 & J1431$-$4715 & J1653$-$2054 & J1729$-$2117\\
		\midrule
		Right Ascension (J2000) & 12:27:00.4413(4) & 14:05:21.4255(8) & 14:31:44.6177(2) & 16:53:31.03(2) & 17:29:10.808(6) \\
		Declination (J2000)     & $-$62:08:43.789(3) & $-$46:56:02.31(1) & $-$47:15:27.574(4) & $-$20:54:55.1(1) & $-$21:17:28(1) \\
		Galactic longitude ($\degree$) & 300.08 & 315.83 & 320.05 & 359.97 & 4.50 \\
		Galactic latitude ($\degree$) & 0.59 & 14.08 & 12.25 & $-$14.26 & $7.22$\\
		\\
        Discovery signal-to-noise & \multirow{2}{*}{11.2} & \multirow{2}{*}{125.0} & \multirow{2}{*}{12.0} & \multirow{2}{*}{11.5} & \multirow{2}{*}{13.7} \\
        ratio & & & & & \\
        Offset from survey& \multirow{2}{*}{$6$} & \multirow{2}{*}{$3$} & \multirow{2}{*}{$1.8$} & \multirow{2}{*}{$2.4$} & \multirow{2}{*}{$4.8$}\\
		beam centre (arcmin) & & & & & \\
		\\
		TOA range (MJD) & 55901--56641 & 55668--56557 & 55756--56627 & 55658--56676 & 55505--56686 \\
		\\
		$P$ (ms)& 34.52783464780(2) & 7.60220343251(1) & 2.0119534425332(9) & 4.129145284562(2) & 66.2928992668(4) \\
        $\dot{P}_\mathrm{meas}$ ($\times 10^{-20}$) & 18.74(7) & 2.79(4) & 1.411(3) & 1.117(6) & 17.2(7) \\
        $\dot{P}_\mathrm{Shk}$ ($\times 10^{-20}$) &   ---   &   2(1) &  0.09(7) & ---  & --- \\
        $\dot{P}_\mathrm{meas} - \dot{P}_\mathrm{shk}$ ($\times 10^{-20}$) & 18.74(7) & 1(1) & 1.32(9) & 1.117(6) & 17.2(7) \\
        DM ($\rm{cm}^{-3}\,\rm{pc}$) & 363.0(2) & 13.884(3) & 59.35(1) & 56.56(2) & 34.49(4)\\
		\\
		DM distance (kpc) & 8.32 & 0.58 & 1.53 & 1.63 & 1.09 \\
		$\rm{S}_{1.4\,\mathrm{GHz}}$ (mJy) & 0.22 & 0.92 & 0.73 & 0.16 & 0.20 \\
		$\rm{L}_{1.4\,\mathrm{GHz}}$ (mJy kpc$^2$) & 15.2 & 0.3 & 1.7 & 0.4 & 0.02 \\
		\\
		$\tau_\mathrm{c}$ ($10^9$ yr) & $2.9$ & $36$ & $2.3$ & 5.9 & 6.1 \\
		$B_\mathrm{surf}$ (G) & $2.6\times 10^{9}$ & $1.6 \times 10^{8}$ & $1.7 \times 10^{8}$ & $2.2 \times 10^{8}$ & $3.4 \times 10^9$ \\
		$B_\mathrm{lc}$ (G) & $5.7 \times 10^2$ & $3.3 \times 10^3$ & $1.9 \times 10^5$ & $2.8 \times 10^4$ & $1.1 \times 10^2$\\
		$\dot{E}$ (erg s$^{-1}$) & $1.8\times 10^{32}$ & $3.0\times 10^{32}$ & $6.8\times 10^{34}$ & $6.3\times 10^{33}$ & $2.3\times 10^{31}$ \\
		$\dot{E}$/$\rm{d}^2$ (erg kpc$^{-2}$ s$^{-1}$) & $2.6\times 10^{30}$ & $8.9\times 10^{32}$ & $2.9\times 10^{34}$ & $2.4\times 10^{33}$ & $1.9\times 10^{31}$ \\
		\\
        Proper motion: & & & & & \\
        $\mu_\mathrm{RA}$ (mas yr$^{-1}$) & --- & $-$44(6) & $-$7(3) & --- & ---\\
        $\mu_\mathrm{Dec}$ (mas yr$^{-1}$) & --- & 20(10) & $-$8(4)  & --- & ---\\
        $\mu_\mathrm{Total}$ (mas yr$^{-1}$) & --- & 48(8) & 11(4) & --- & ---\\
        \\
		Binary Model & ELL1H & ELL1 & ELL1 & ELL1 & --- \\
		\\
		Orbital Period (d) & 6.721013337(4) & 8.95641988(7) & 0.4497391377(7) & 1.226815259(9) & ---\\
		$a\sin{i}$ (lt-s) & 23.200663(3) & 6.567659(9) & 0.550061(2) & 0.688855(6) & --- \\
        TASC (MJD) & 55991.1937918(2) & 55132.23096(2) & 55756.1047771(4) & 55584.728649(3) & --- \\
        T0 (MJD) & 55991.7000(2)  & 55694.0(4) & 55756.23(2) & 55584.9(5) & --- \\
        $\epsilon_1$ & 0.0005238(3) & 0.000005(2) & 0.000023(8) & 0.00000(2) & ---\\
        $\epsilon_2$ & 0.0010229(3) & 0.000004(2) & $-0.000003(7)$ & 0.00000(2) & ---\\
		$e$ & 0.0011494(3) & 0.000007(2) & 0.000023(8) & 0.00001(3) & --- \\
		$\omega$ (deg) & 27.11(1) & 51(16) & 97(18) & 0(30) & --- \\
		\\
        \multicolumn{6}{l}{Shapiro Delay Parameters} \\
        $h_3$ ($\times 10^{-6}$) & $7.3 \pm 2.7$  & --- & --- & --- & --- \\
        $h_4$ ($\times 10^{-6}$)& $4.4 \pm 3.2$  & --- & --- & --- & --- \\
        \\
		Min. $m_\mathrm{c}$ ($\mathrm{M}_\odot$) & 1.27 & 0.21 & 0.12 & 0.08 & --- \\
		Med. $m_\mathrm{c}$ ($\mathrm{M}_\odot$) & 1.58 & 0.25 & 0.14 & 0.09 &  ---\\\\
		RMS of fit ($\upmu$s)                     & 22 & 26 & 10 & 35 & 162 \\
        Reduced $\chi^2$ & 1.30 & 1.05 & 1.8$^{*}$ & 1.7 & 2.7 \\
		\bottomrule
        \multicolumn{6}{l}{\footnotesize $^{*}$When TOAs around superior conjunction are removed}\\
		\end{tabular}
		\label{fullsolns}
\end{table*}

\section{Discovery and Timing}
\label{s:disc_timing}

The five pulsars presented here were discovered in the HTRU survey for pulsars 
and transients \citep[for a full discussion, see][]{kjv+10}.
Survey observations were made with the 64-m 
Parkes radio telescope using the 13-beam $21\text{-cm}$ multibeam receiver.  
The 
survey has an observing band centred at $1352 \text{ MHz}$ and has a useful bandwidth
 of $340\text{ MHz}$.  The HTRU survey is split into three areas: the low-, mid- and 
high-Galactic latitude regions.  The discoveries presented here are from the 
mid-latitude survey, which tiles $ -120\degree < l < 30\degree$ and
$|b| \leq 15\degree$ with observations of $540\mathrm{~s}$.

The data were processed using the Fourier transform based pulsar search pipeline 
described in \citet{kjv+10}.  The discoveries presented 
here were initially identified using an artificial neural network which aims
 to highlight the best candidates from the survey for human inspection 
\citep{emk+10,bbb+12}.  After inspection, confirmation observations were taken at
 the sky positions of the survey beams deemed to contain the best candidates.  
These confirmations were performed with the Parkes or Lovell telescopes.
  Four of the five pulsars presented here, PSRs 
J1227$-$6208, J1405$-$4656, J1431$-$4715, and J1653$-$2054, are in orbit with a 
binary companion,  while one, PSR J1729$-$2117, is isolated. 
PSR J1227$-$6208 was also independently discovered in the PMPS by two separate groups
\citep{mlb+12,kek+13}. All observations, timing, and analysis of 
PSR J1227$-$6208 presented here are independent work. 

After confirmation, the new pulsars were timed regularly with Digital Filterbank
backend systems; PSRs J1653$-$2054 and  J1729$-$2117 with the 76-m Lovell
 telescope at Jodrell Bank Observatory (JBO) and PSRs J1431$-$4715, J1405$-$4656,
 and J1227$-$6208 with the 64-m radio telescope at Parkes (see Table \ref{table:timing}).
  The pulsars timed with the Lovell telescope were observed approximately once per
fortnight, whereas those observed using the Parkes radio telescope were observed more
 sporadically, while maintaining phase coherence, with one case of an approximately
 80 day gap between observations.

Each observation resulted in a pulse time-of-arrival (TOA) measurement.  
 Parameters were fitted to the TOAs using the \textsc{tempo2} software 
 package \citep{hem06} 
and the best fit parameters for 
the five recycled pulsars are given in Table~\ref{fullsolns}.

\section{Results}
\label{s:results}

The pulse periods for the new discoveries range
from $2.01 \text{ ms}$ for PSR J1431$-$4715, 
placing it among the 20 fastest spinning pulsars, to $66.29 \text{ ms}$ 
for PSR J1729$-$2117, one of the longer periods for a partially
recycled pulsar \citep{mhth05}\footnote{www.atnf.csiro.au/people/pulsar/psrcat}.
 One, PSR J1729$-$2117, has not exhibited any detectable
 periodic variation of the pulse period over the 3 years of timing, indicating that 
 it is one of just two isolated recycled pulsars discovered in the HTRU
 survey to date.  The five pulsars represent some of the different types of 
 known recycled pulsar and binary systems.

 \begin{figure}
	\begin{center}
	\includegraphics[width=8.5cm]{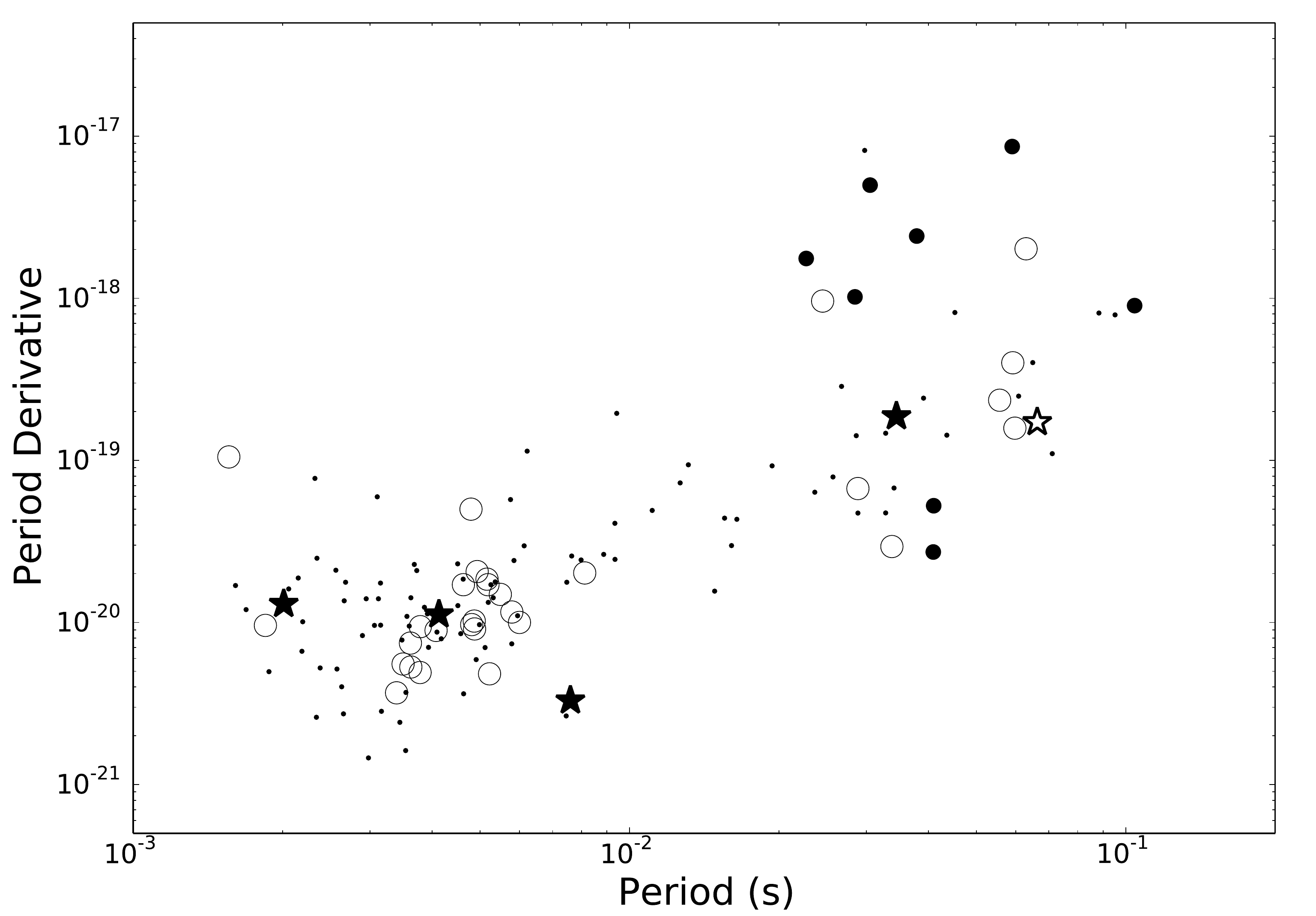}
	\end{center}
	\caption{A plot of pulse period 
    derivative, $\dot{P}$, against pulse period, $P$, using the intrinsic values of $\dot{P}$ given in Table 1.  Double neutron star (DNS) systems (see Table \ref{t:dns}) are plotted as large filled circles, other binaries are small filled circles, and isolated systems are unfilled circles. The discoveries presented in this work are shown as stars, with the isolated system PSR~J1729$-$2117 an unfilled star. Only non-globular cluster pulsars with $P < 100 \text{ ms}$ and $\dot{P} < 10^{-16}$ and DNS systems are shown.} 
	\label{f:ppdot_msp}
\end{figure}

\begin{figure*}
	\begin{center}
	\includegraphics[width=16cm]{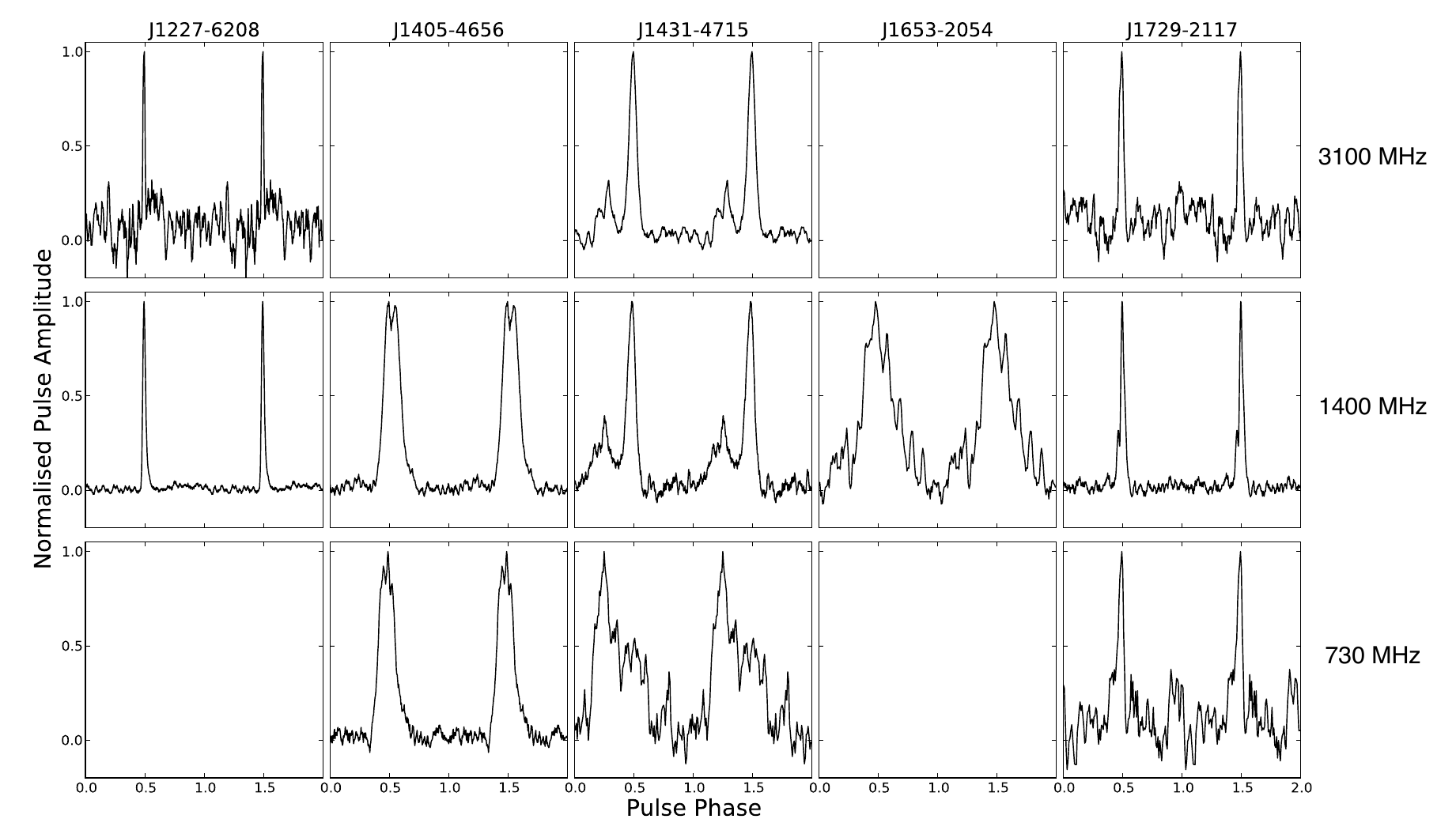}
	\end{center}
	\caption[Multi-frequency MSP pulse profiles]{Typical pulse profiles are shown 
        for the five millisecond pulsars presented here, from single observations. Profiles have been shifted so that the peaks of the 1400~MHz profile sits at a pulse phase of 0.5. Top, middle, and 
bottom rows correspond to observing frequencies of $3100\:\text{MHz}$, $1400\:\text{MHz}$, and $730\:\text{MHz}$,
 respectively.  Blank plots correspond to non-detections. The observing times are
 given in Table \ref{table:timing}.  All observations are taken with the 64-m 
Parkes radio telescope, except for the 1.4 GHz observations and timing of PSRs 
J1653$-$2054 and J1729$-$2117, which use the 76-m Lovell telescope at Jodrell Bank. }
	\label{f:profiles}
\end{figure*}

\subsection{Period Derivatives}

The measured spin-down rate, $\dot{P}_\mathrm{meas}$, of a pulsar with period $P$ can 
differ from the intrinsic spin-down via the Shklovskii effect \citep{shklovsky1970}. 
A proper motion of $\mu$, for a pulsar with period $P$ at distance 
$d$, leads to an extra contribution to the period derivative as
\begin{equation}
    \dot{P}_\mathrm{Shk} \simeq 2.43 \times 10^{-21} \left( \frac{P}{\mathrm{s}}\right) \times 
        \left( \frac{\mu^2}{\mathrm{mas~yr}^{-1}} \right) \times 
        \left( \frac{d}{\mathrm{kpc}} \right),
\label{shklovsky}
\end{equation}
where $c$ is in units of $\mathrm{m\,s}^{-1}$,
which can be a considerable contribution for some MSPs since their rotational
periods are so short, the measured $\dot{P}$ so small, and transverse velocities 
are typically $\sim 85~\mathrm{km\,s}^{-1}$ \citep{toscano1999}. This being the case,
the Shklovskii effect is likely to give only small contributions (of the order 
$\lesssim 10\%$) to the measured 
$\dot{P}$ for PSRs~J1227$-$6208 and J1653$-$2054 (for which no significant
proper motion has been measured), but for 
PSR~J1729$-$2117, the Shklovskii effect could easily contribute
a significant fraction, $\gtrsim 30\%$, of the measured $\dot{P}$. 


In the cases of PSRs~J1405$-$4656 and J1431$-$4715, which have measured, significant,
proper motions of 48 and 11~mas~yr$^{-1}$, respectively. The Shklovskii effect then 
contributes $\sim 90$\% and $\sim 5$\% of the measured $\dot{P}$ for these pulsars.
In Table~\ref{fullsolns} we include corrections for the Shklovskii effect, where 
it may be calculated. However, the errors on the value of $\dot{P}_\mathrm{Shk}$
(and, hence, the intrinsic period derivative)
are rather large, especially in the case of PSR~J1405$-$4656, 
due mainly to contributions from the error on the proper motion
and the distance derived from the electron distribution model, 
which we have assumed to be 30\%.

\subsection{PSR J1227$-$6208}

PSR J1227$-$6208 is located in the region of 
the $P$-$\dot{P}$ diagram where mildly recycled pulsars are found (see
 Figure \ref{f:ppdot_msp}).  This is consistent with it having been spun-up during
 unstable, short-lived mass transfer in an HMXB phase. Pulse profiles from 
 observations at 1.4 and 3.1~GHz are shown in Figure~\ref{f:profiles}.

It has an orbital eccentricity, $e = 0.00115$, 
orbital period $P_{\text{orb}} = 6.72 \text{ days}$, and a projected semi-major axis, 
$a_{\text{p}}\sin (i) = 23.2 \text{ ls}$.  
Assuming $\text{M}_{\text{p}} = 1.4 \text{ M}_{\odot}$, this leads to a minimum companion
mass of $\text{M}_\text{c, min} = 1.29 \text{ M}_\odot$ for an edge-on orbit.
  The low eccentricity and high minimum companion mass are unusual: there are only
  three other systems known with $e < 0.1$ and $\text{M}_{\text{c, min}} > 1.0\text{ M}_{\odot}$.  

PSR J1227$-$6208 is similar to other so-called Intermediate Mass Binary Pulsars 
(IMBPs); for example PSR J1435$-$6100 \citep{clm+01} and PSR J2222$-$0137 
\citep{blr+13}.  PSR J1227$-$6208's spin parameters are also similar to 
PSR J0609$+$2130 --- an isolated pulsar thought to be the result of an HMXB 
which disrupted during the second ccSN \citep{lma+04}. In an IMBP it is thought
 the companion was not sufficiently massive to undergo a ccSN, resulting instead 
in the formation of a heavy CO or ONeMg WD.  The minimum mass for the companion 
to PSR J1227$-$6208 is common for ONeMg WDs, which have a mass range around 
$1.1$-$1.3\text{ M}_{\odot}$ \citep{tlk12}.  This is, however, the \emph{minimum} 
mass the companion may have; it is already close to the upper limit of ONeMg WDs,
 and to the Chandrasekhar mass limit. 

For inclinations $i < 71\degree$, $\text{M}_\text{c} \gtrsim 1.4 \text{ M}_\odot$, and
 therefore PSR~J1227$-$6208 might be part of a DNS system.  If the companion is a NS,
though not one which is visible to date as a pulsar, it would have formed from a 
massive progenitor ($8$-$10 \text{ M}_{\odot}$), and the system would have evolved
 through an HMXB phase, hence leaving PSR~J1227$-$6208 partially recycled. 
 Compared to the known DNS systems, listed in Table \ref{t:dns},
 the eccentricity is small, and the formation of DNS systems with such a small 
eccentricity appears to be unlikely \citep{cb05}.
  Simulations by \citet{dpp05} and \citet{cb05} do, however, suggest 
that such low-eccentricity DNS systems could exist because of the possibility
 of very small and retrograde ccSN kinematic kicks, although they would be extremely rare.

 {\bf Due to the loss of binding energy ($\delta M \gtrsim 0.1$M$_\odot$)
during a SN explosion, the mass loss induces a finite eccentricity,
$e = (M + \delta M)/M - 1 \gtrsim 0.035$, into the system, which is 
$\sim 30$ times that of PSR~J1227$-$6208. The only way this can be lessened
is if the exploding star receives a retrograde kick into an exceptionally small
volume of phase space to reduce the angular momentum and, hence, eccentricity
\citep{cb05}. For neutron stars receiving a random kick from a large kick velocity
distribution the probability is vanishingly small ($<$0.1\%), but the double pulsar's
eccentricity of 0.09 \citep{lbk+04} demonstrates that not all neutron stars receive large
kicks. Even if the kick was constrained to be small and confined to the pre-SN
orbital plane, PSR~J1227$-$6208's eccentricity of 0.001 would occur less than 
$e_\mathrm{kick}/e_\mathrm{psr} \sim 3\%$ of the time as most random kicks
act to increase, not decrease, the orbital eccentrity induced by mass loss.
On the other hand, such low-eccentricity systems would not lose energy via 
gravitational wave emission as quickly as eccentric systems, and consequently 
will survive longer before merging.}

\begin{table}
	\begin{center}
	\caption{Basic orbital and spin parameters
 for known DNS systems.  Also included are the same parameters for PSR J1227$-$6208.}
		\begin{tabular}{lccccr}
		\hline
		Pulsar&$P$ & $P_{\text{orb}}$ &  $a_{\text{p}}\sin i$ &  $e$  & $\text{M}_{\text{c}}$ \\
        & (ms) & (d) & (ls) & & ($\text{M}_{\odot}$) \\
		\hline
J0737$-$3039 & $22.69$ & $0.102$ & $1.415$ &  $0.088$ &$1.250$ \\
J1518$+$4904 & $40.93$ & $8.634$ & $20.044$ & $0.249$ & $> 0.83$ \\
B1534$+$12 & $37.90$ & $0.420$ & $3.729$ & $0.273$ & $1.34$ \\
J1756$-$2251 & $28.46$ & $0.319$ & $2.756$ & $0.180$ & $1.18$ \\ 
J1811$-$1736 & $104.18$ & $18.779$ & $34.783$ & $0.828$ & $>0.87$ \\
J1829$+$2456 & $41.00$ & $1.176$ & $7.236$ & $0.139$ & $>1.22$ \\
B1913$+$16 & $59.03$ & $0.322$ & $2.341$ & $0.617$ & $1.38$\\
B2127$+$11C & $30.52$ & $0.335$ & $2.518$ & $0.681$ & $1.354$ \\\\
J1227$-$6208 & $35.52$ & $6.721$ & $23.200$ & $0.001$ & $> 1.29$ \\ 
		\hline
		\end{tabular}
		\label{t:dns}
	\end{center}
\end{table}

With $\text{DM} = 362.6\text{ cm}^{-3}\text{ pc}$, the NE2001 
electron density model predicts  PSR~J1227$-$6208 to be located at 
a distance of $8.3\text{ kpc}$.  This would make an optical 
detection of a WD companion (for the case of a large orbital inclination)
extremely difficult.   Consequently, the best chance 
of measuring the masses in the system comes from the potential measurement of 
post-Keplerian parameters. 

\subsubsection{Periastron advance}
\label{sec:peri}
It is possible to predict relativistic 
changes to the orbit due to general relativity.   One of the 
easiest post-Keplerian parameters to measure for systems with eccentric orbits
is a changing longitude of periastron,

\begin{equation}
    \dot{\omega} \simeq 39.73\degree \mathrm{~yr}^{-1} \left( \frac{P_{\text{orb}}}{\mathrm{hr}} \right)^{-5/3}\left(\frac{1}{1-e^2}\right) \left(\frac{M_{\text{c}} + M_{\text{p}}}{\mathrm{M}_\odot}\right)^{2/3}
\label{e:omdot2}
\end{equation}

\noindent where $P_{\text{orb}}$ is
the orbital period, $e$ is the orbital eccentricity, and $M_{\text{c}}$ and 
$M_{\text{p}}$ are the companion and pulsar masses respectively.  A measured 
$\dot{\omega}$ constitutes a measurement of the combined masses,
 $M_{\text{c}} + M_{\text{p}}$ and consequently constrains the companion mass. 
 While the orbital eccentricity of PSR J1227$-$6208 is small, it is 
 significantly higher than is typical for systems which have likely evolved
 through an LMXB phase \citep{phi92}.
 Equation \ref{e:omdot2} predicts $\dot{\omega} = 0.017\degree \text{~yr}^{-1}$
 for the PSR J1227$-$6208 system if we assume pulsar and companion masses
 of $1.4 \text{~M}_\odot$. For a low inclination, and
 $M_{\text{c}} = 10 \mathrm{~M}_\odot$, we would then expect 
$\dot{\omega} = 0.042\degree \text{~yr}^{-1}$.

Current observations have not made a significant measurement of $\dot{\omega}$, 
however, we can make some quantitative statements about the value of $\dot{\omega}$.
Using the \textsc{fake} plugin for \textsc{tempo2}, TOAs were generated for 
PSR~J1227$-$6208 using the parameters listed in Table~\ref{fullsolns}, varying
$\dot{\omega}$ from 0.01 to 0.05$\degree \mathrm{~yr}^{-1}$. 
Fitting those TOAs using 
\textsc{tempo2} over the current data span, $\dot{\omega}$ was only measured 
at the 2-$\sigma$ level for $\dot{\omega} \gtrsim 0.04\degree \mathrm{~yr}^{-1}$. 
Since we are unable to make a significant measurement, it is likely 
that $\dot{\omega} \leq 0.04\degree \mathrm{~yr}^{-1}$.
Extending these simulated TOAs into the future, in all cases another two years of 
data constrain $\dot{\omega}$ at the $\sim 3$-$\sigma$ level, which would allow
a direct measurement of $M_{\text{p}}$ and $M_{\text{c}}$ at the 2-$\sigma$ level.

\subsubsection{Shapiro delay}
\label{ss:shap}
The radio pulses from a pulsar in a binary system are delayed as they cross 
the gravitational potential of the companion in a phenomenon called 
Shapiro delay \citep{sha64}. The magnitude of the delay,
$\Delta_{\text{S}}$, is quantified by two post-Keplerian 
parameters, the range, $r$, and the shape, $s$. For low-eccentricity systems,
\begin{equation}
\Delta_\text{S}(\Phi) = -2r \text{ln}(1-s\sin \Phi) ,
\label{e:shap}
\end{equation} 

\noindent where $\Phi$ is the orbital phase ($\Phi = 0$ is the ascending node),
$r = \text{T}_{\odot}\text{M}_{\text{c}}$, $s = \sin i$, and
$\text{T}_{\odot} = \text{GM}_{\odot}/c^3$.  The Shapiro delay is the
 same for every orbit of the system and so when the TOA residuals are folded 
modulo-$P_{\text{orb}}$ a characteristic shape in the timing residuals may be 
measured.

Since the companion mass for PSR~J1227$-$6208 is relatively high, implying a large
value of $r$, an observing programme was undertaken using the 64-m Parkes radio
telescope to obtain TOAs across the expected epoch of superior conjunction,  
and attempt to observe Shapiro delay in this system.
A global timing solution was made (given in
Table \ref{fullsolns}) using the parameterisation for the Shapiro delay 
given by \citet{fw2010}, where the fitted parameters $h_3$ and $h_4$ are
related to the Shapiro $r$ and $s$ parameters as
\begin{equation}
    r = \frac{h_3^4}{h_4^3} , ~~s = \frac{2h_3 h_4}{h_3^2 + h_4^2}.
\label{e:h3h4}
\end{equation}
A best-fit value of $h_3$ which is consistent with zero implies a non-detection of the 
Shapiro delay, and a measurement of $h_4$ allows the orbital inclination
to be constrained.

\begin{figure}
	\begin{center}
	\includegraphics[width=8.5cm]{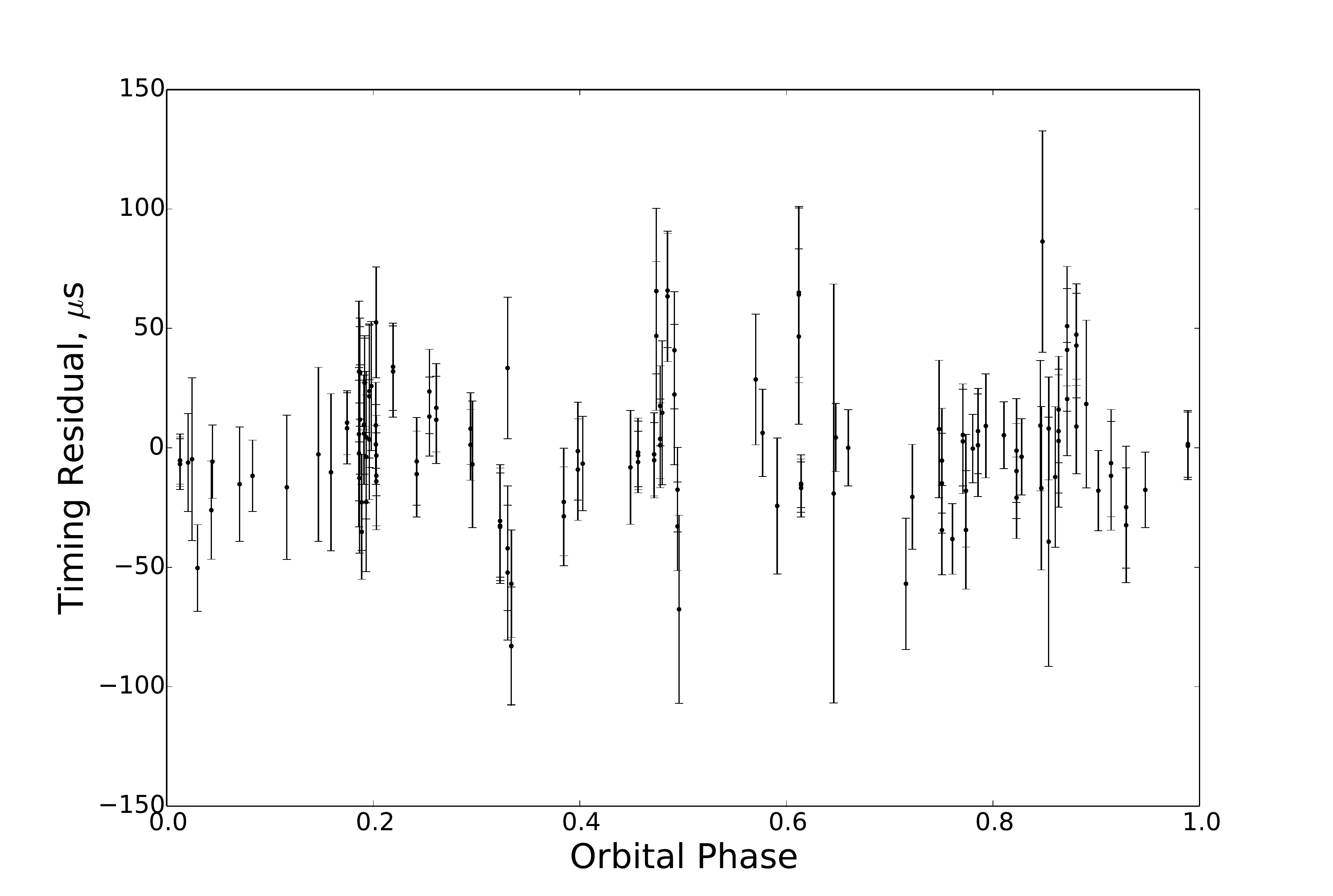}
	\end{center}
    \caption{Timing residuals from the best fit global model for PSR~J1227$-$6208 as a function of orbital phase, without any Shapiro delay contribution taken into account. Although the timing residuals appear to show some structure, fitting for the Shapiro delay provides only marginal measurements of orbital parameters.}
	\label{f:shap_toas}
\end{figure}

Timing residuals for PSR~J1227$-$6208, shown in Figure~\ref{f:shap_toas} were 
observed to show some evidence of Shapiro delay due to the apparent non-Gaussianity
of the residuals. Fitting for $h_3$ and $h_4$, we obtained best fit values of
$h_3 = (7.3 \pm 2.7) \times 10^{-6}$
and $h_4 = (4.4 \pm 3.2) \times 10^{-6}$; however both of these measurements are
to a low significance, and corresponding errors in $M_\mathrm{c}$ and $\sin i$
are large. Nevertheless, the resulting values of range and shape are
$r = (3.3 \pm 8.8) \times 10^{-5} \mathrm{~M_\odot s}$ and 
$s = \sin i = 0.88^{+0.12}_{-0.33}$.

To estimate when the Shapiro delay might become well-constrained, as in 
\S~\ref{sec:peri}, TOAs were generated for PSR~J1227$-$6208 using the 
\textsc{Tempo2} plugin \textsc{fake}. Keeping the RMS fixed at 
22~$\upmu$s, a further 9 years of 
timing data were needed in order to constrain both $h_3$ and $h_4$ at the 
3-$\sigma$ level. Alternatively, using the current data span, $h_3$ and $h_4$ 
were constrained at the 3-$\sigma$ level when the RMS 
was reduced to 10~$\upmu$s. Therefore, a combination of improved timing and a 
longer data span may enable precise measurement of $h_3$ and $h_4$ in under 9 years.

%

\subsection{PSR J1431$-$4715}

PSR J1431$-$4715 has the 14th shortest spin period of all known pulsars, 
$P = 2.01\text{ ms}$, and a small period derivative, 
$\dot{P} = 1.4 \times 10^{-20}$.  These values place 
PSR J1431$-$4715 firmly amongst the fully recycled MSPs in the bottom left of
 the $P$-$\dot{P}$ diagram (see Figure \ref{f:ppdot_msp}).  
 
Observations of PSR J1431$-$4715 at multiple frequencies (see Figure \ref{f:profiles})
reveal significant 
pulse profile evolution with observing frequency, as well as evidence of 
pulse delays and eclipses when the pulsar is at superior conjunction. 
At $1.4\text{ GHz}$ the pulse
 profile is double peaked, with a smaller and wider leading component, the 
profile at $3.1\text{ GHz}$ is similar although the leading component appears 
to be weaker.  At $0.732\text{ GHz}$ (observed away from superior conjunction) 
the trailing component has 
become the larger of the two. Strong pulse profile evolution has also been measured 
in other eclipsing systems, such as PSR J2215$+$5135 \citep{hrm11}.  
Further observations of PSR J1431$-$4715 at a range of frequencies will prove 
useful in both further study of the pulse profile evolution and 
measuring the spectral index.

\begin{figure}
	\includegraphics[width=8.5cm]{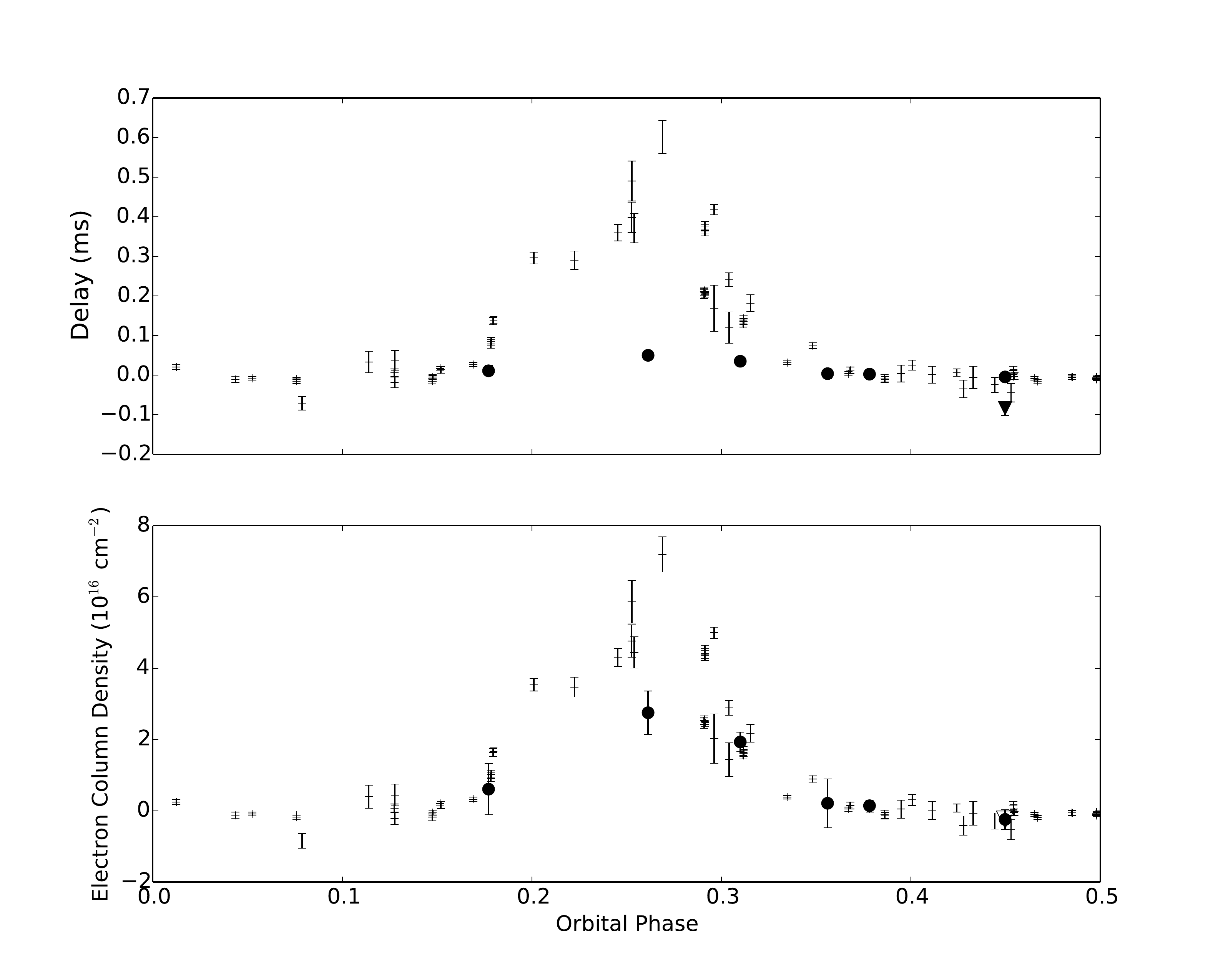}

    \caption[Residual TOAs for PSR J1431$-$4715]{The top panel shows the timing residuals for PSR~J1431$-$4715 which are folded modulo-$P_{\text{orb}}$, between orbital phases of 0 and 0.5.  The excess residuals which are not described by the timing model are centred around superior conjunction, an orbital phase of 0.25.  The lower panel shows the same delays converted to an electron column density within the eclipsing region. In both panels there are observations at three frequencies; 0.732 GHz (triangle), 1.4 GHz (dots), and 3.1 GHz (circles).}  
	\label{f:1431res}
\end{figure}

This pulsar also 
exhibits significant orbital-phase-dependent delays in 
the pulse TOAs near superior conjunction (Figure~\ref{f:1431res}).
 The delays can be explained by an excess DM, attributed
 to the passage of the radio pulses through ionised material surrounding the 
companion. The timing model described in Table 
\ref{fullsolns} was generated by excluding TOAs which are obviously 
associated with the eclipse region, $0.1<\Phi <0.4$.

Multi-frequency observations indicate that the magnitude of the eclipse delays
depends upon observing frequency, with the $3.1\text{ GHz}$ observations
 showing somewhat shorter delays than at $1.4\text{ GHz}$. 
 Since this difference is removed when we convert from time delays to additional
 electrons along the line of sight, we can conclude that these delays are purely
 due to additional DM introduced by the material between the pulsar and Earth.
 The lower frequency
 observations (centred on $0.732\text{ GHz}$) have not resulted in a positive 
detection to date across the eclipse region (see Figure \ref{f:1431res}). Treating
the delay as purely dispersive, we can relate the delay time, $t$,
to the DM as
\begin{equation}
    t = \frac{e^2}{2 \pi m_e c} \frac{\mathrm{DM}}{f^2}
\end{equation}
for observations at a frequency $f$. 
In Figure~\ref{f:1431res}, this DM has then been
converted to the free-electron column density by approximating the depth of the 
eclipsing region as being equal to the radius of the eclipse region;
the computed electron column density is similar to other eclipsing systems; 
for example PSRs J2051$-$0827 \citep{sbl+96} and J1731$-$1847 \citep{bbb+11}.

Using multiple observations during different eclipses we find that there appear
 to be significantly different delays at the same orbital phase within the eclipse
 region.  The TOAs were measured during observations which were separated by 
several months, that is $\sim10^2$ orbital periods.  As such, the amount, or density,
 of dispersive material must be variable on this timescale.  The width of the 
eclipsing region appears to be constant, suggesting it is the free electron density
 which is variable as opposed to depth of the eclipsing region. 

\begin{figure}
	\begin{center}
	\includegraphics[width=8.5cm]{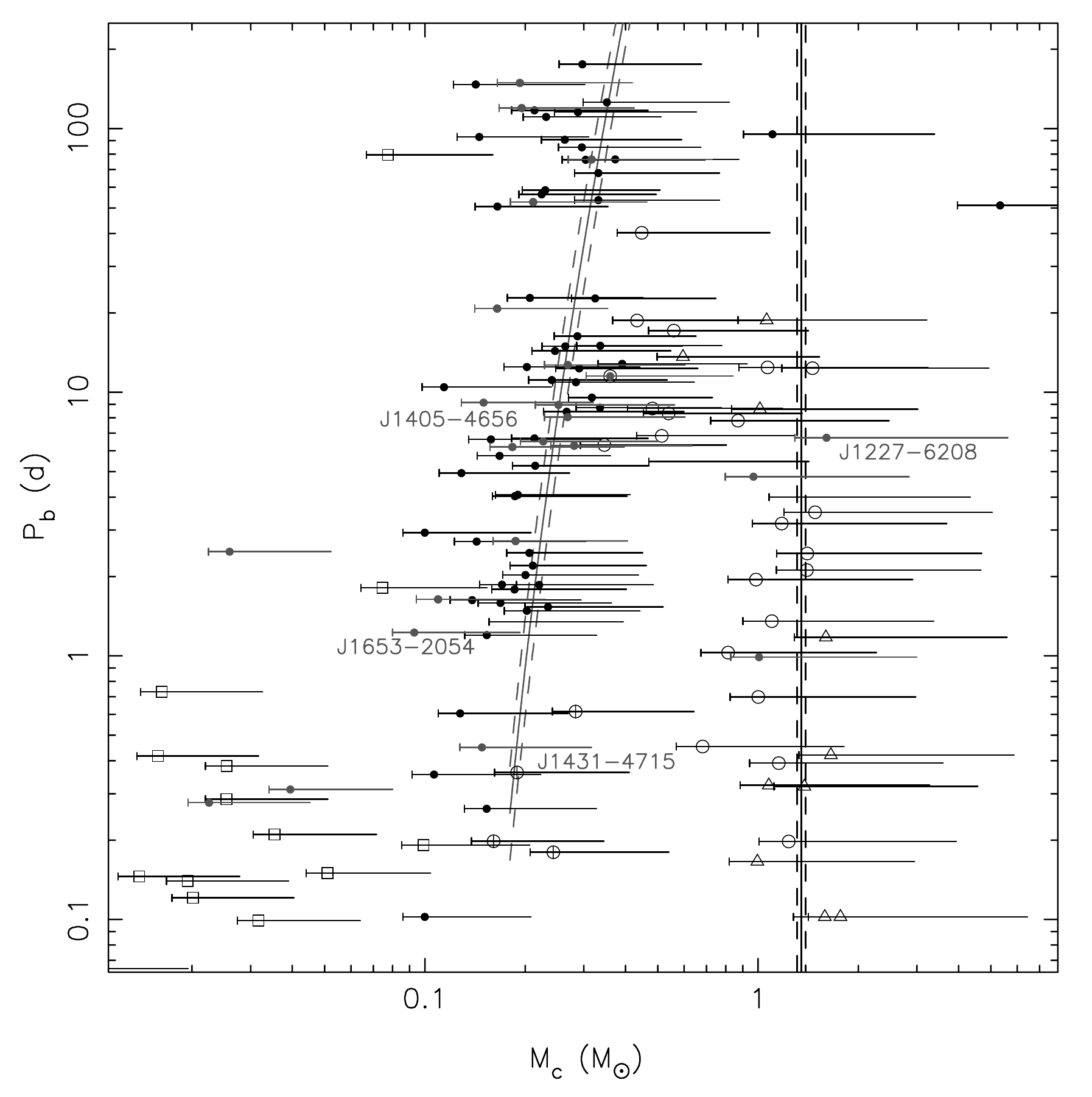}
	\end{center}
    \caption{A plot of orbital period against companion mass for non-globular cluster pulsars.  Data points correspond to the median mass value, while lower and upper limits correspond to the minimum companion mass and the 90\% confidence upper limit.  The four binary systems presented in this paper are marked with text, and the other symbols correspond to the systems presented in Table~4.
    The curves indicate the relationship for HeWDs derived by \citet{ts99a} for three different companion progenitors, and the vertical lines denote a companion mass of 1.4~M$_\odot$, 
indicative of a double neutron star system.}  
	\label{f:ts99}
\end{figure}

From the orbital separation  and the mass ratio we determine the approximate 
distance of the Roche lobe from the companion, $R_L$, using 

\begin{equation}
R_{\text{L}} = \frac{0.49 A q^{2/3}}{0.6q^{2/3} + \text{ln}\left( 1 + q^{1/3} \right) } ,
\label{e:rl}
\end{equation}

\noindent from \cite{egg83}, where $A$ is the separation of the pulsar and its companion, 
and $q = m_\mathrm{c} / m_\mathrm{p}$ is the mass ratio of the system. 
In this case $R_{\text{L}} = 0.6 \text{ R}_{\odot}$,
 about half the size of the eclipse radius. 
 A significant amount of the material in the eclipsing region
 is therefore outside the companion's Roche lobe. 
Material which is close to the Roche lobe can easily be removed by a relativistic
 pulsar wind \citep{bkr+13}.  This would mean some eclipsing material was not 
gravitationally bound to the companion and must therefore be continually 
replenished \citep{sbl+96}. 

The spin-down dipole radiation $\dot{E}$ of the pulsar at the distance of the 
companion is $\dot{E}/A^2 \sim{} 1.5 \times 10^{33} \text{ ergs s}^{-1}\text{ ls}^{-2}$.   This is typical of other eclipsing systems in general and in the HTRU sample
 \citep{rob11,bkr+13}.  Incident spin down energy at the companion is  
significantly larger for eclipsing systems than non-eclipsing binaries, 
 an indication that the pulsar's energy output is at least partly responsible 
for the dispersive material surrounding the companion.

\begin{table*}
	\begin{center}
		\begin{tabular}{lcccrcrrcc}
		\hline
        Name & $d$ & $\tau_\mathrm{c}$ & $M_\mathrm{c, med}$ & $A_V$ & $(m-M)_V$ & $M_V$ & $m_V$ & Reference\\
             & (kpc) & (Gyr) & & & & & & & \\
    \hline
    J1017--7156 & 2.87 & 16.69 & 0.226 &  0.263 & 12.55 & $>13$ & $>25.5$ & \citet{kjv+10} \\
    J1056--7117 & 2.60 &  6.60 & 0.150 &  0.388 & 12.46 & $<10$ & $<22.5$ & \citet{ngetal2014} \\
    J1125--5825 & 2.58 &  0.81 & 0.317 &  1.765 & 13.83 & $<13$ & $<26.8$ & \citet{bbb+11} \\
    J1405--4656 & 0.58 & 37.64 & 0.252 &  0.885 &  9.70 & $>16$ & $>25.7$ & This paper \\ 
    J1431--4715 & 1.53 &  2.26 & 0.148 &  2.290 & 13.21 & $>10$ & $>23.2$ & This paper \\
    \\
    J1431--5740 & 2.52 & 10.14 & 0.187 &  2.821 & 14.83 & 12--16 & $>26.8$ & \citet{burgay2013} \\
    J1528--3828 & 2.20 &  5.00 & 0.195 &  1.447 & 13.16 & 11--16 & $>24.2$ & \citet{ngetal2014}\\
    J1543--5149 & 2.43 &  2.02 & 0.268 &  4.413 & 16.34 & 10--14 & $>26.3$ & \citet{kjv+10} \\
    J1545--4550 & 2.10 &  1.08 & 0.183 &  2.580 & 14.19 & 10--12 & $>24.2$ & \citet{burgay2013} \\
    J1708--3506 & 2.77 &  6.25 & 0.193 &  2.534 & 14.75 & 11--16 & $>25.8$ & \citet{bbb+11} \\
    \\
    J1755--3716 & 3.90 &  6.40 & 0.360 &  0.400 & 13.36 & 11--16 & $>24.4$ & \citet{ngetal2014}\\
    J1801--3210 & 3.95 & 44.60 & 0.165 &  0.687 & 13.67 & $>10$ & $>23.7$  & \citet{bbb+11} \\
    J1811--2405 & 1.77 &  3.15 & 0.280 & 11.360 & 22.60 & 11--15 & $>33.6$ & \citet{bbb+11} \\
    J1825--0319 & 3.03 & 10.61 & 0.211 &  9.210 & 21.62 & $>12$ & $>33.8$  & \citet{burgay2013} \\
    J2236--5527 & 0.83 & 11.40 & 0.268 &  0.063 &  9.66 & 12--16 & $>21.7$ & \citet{burgay2013} \\
\hline
		\end{tabular}
	\end{center}
    \caption{Predicted optical magnitudes for the HTRU MSPs with helium-core white dwarf companions. Limits are calculated using a lower limit on the pulsar spin-down age of $0.1\tau_\mathrm{c}.$}
	\label{t:optparams}
\end{table*}

\subsection{PSR J1653$-$2054}

PSR J1653$-$2054 is a fully recycled MSP with $P = 4.129\text{ ms}$ and
 $\dot{P} = 1.14 \times 10^{-20}$. The pulse profile at 1.4~GHz is
 shown in Figure~\ref{f:profiles}. The minimum mass of the companion to 
PSR J1653$-$2054 is just $0.08 \text{ M}_{\odot}$, placing it between typical 
black widow and redback systems.  The measurement of 
$\dot{E}/A^2 \sim{} 3.7 \times 10^{31} \text{ ergs s}^{-1}\text{ ls}^{-2}$ 
is however somewhat lower than for the redback and black widow systems 
(and PSR J1431$-$4715).  It is, therefore, possible that the incident pulsar spin-down
 energy at the surface of the companion is too low to bloat the companion, which 
 would explain why no eclipses have been observed. The orbital period is at least a 
factor of two longer than known Galactic redback and black widow systems \citep{rob11}.
  The wide orbit means that any ionised region would subtend a
 smaller angle, making eclipses less likely.  If, however, 
the orbit is not being viewed edge-on then the companion mass would be higher, 
for instance, if $i < 25\degree$ then $\text{M}_\text{c} > 0.2 \text{ M}_{\odot}$, 
indicative of a typical HeWD. A detection, or constraint on the magnitude, 
of the optical companion to this system may help us resolve the nature 
of the companion and whether this is an example of a wide
orbit black widow or redback.

\subsection{PSR J1729$-$2117}
This system, with period $P = 66.29\text{ ms}$ and period derivative 
$\dot{P} = 1.6\times 10^{-19}$,
has very similar spin parameters to PSR J0609$+$2130 \citep{lma+04}
 and PSR J2235$+$1506 \citep{cnt93}. \citeauthor{cnt93} suggested that PSR J2235$+$1506
 was spun-up in an HMXB system in which the second ccSN resulted in the disruption 
of the binary.  It may be that similar arguments can be applied to PSR J1729$-$2117. 
As discussed by \citet{burgay2013}, this pulsar was only detected owing to 
an enhanced flux density due to scintillation during the discovery observation.
Subsequent observations have revealed it to have a low luminosity (Table 2).

If this pulsar really was spun up in a binary, then PSR J1729$-$2117 
might be expected to have a high velocity
 \citep{bai89}, and as the system is relatively close (the distance from
 NE2001 is $d = 1.09\text{ kpc}$), the proper motion should be measurable with 
 further timing. Pulse profiles at 0.7, 1.4 and 3.1~GHz are shown in 
 Figure~\ref{f:profiles}.

\subsection{Orbital periods and companion masses}
The five MSPs presented here show quite diverse orbital properties. One, 
PSR~1729$-$2117, appears to be isolated and has presumably lost its companion
at some point since it was recycled. Three pulsars, PSRs~J1405$-$4656, 
J1431$-$4715, and J1653$-$2054, all have low to intermediate mass companions 
of $\sim 0.1$ to 0.2$\mathrm{ M}_\odot$, and finally PSR~J1227$-$6208 has 
a high-mass companion of $\gtrsim 1.3\mathrm{ M}_\odot$.

\cite{ts99a} derived a relationship between orbital period and companion mass 
for binary MSPs with WD companions which have formed via an LMXB phase (see 
Figure \ref{f:ts99}).  This relationship is based on a predictable core mass for
 a main sequence star as a function of stellar radius.  During spin-up in an 
LMXB, the edge of the star is at the position of the Roche lobe, which is a 
function of only the two masses and the orbital separation (see Equation \ref{e:rl}).
  After the companion star exhausts its fuel and the outer layers are blown off, 
the stellar core becomes a WD.  The WD mass is therefore linked to the orbital 
period at cessation of mass transfer/spin-up.  

Of the binary systems described here, PSR J1653$-$2054 has a shorter orbital period
 than those described by \cite{ts99a}, who only considered 
$P_{\text{orb}} \gtrsim 2 \text{ days}$.  PSRs J1405$-$4656 and J1431$-$4715 are
 in agreement with this relationship, and since PSR J1227$-$6208 did not form via 
 an LMXB phase, on which this relationship is based, it does not lie on the 
 predicted curve. Indeed, its minimum companion mass is considerably higher
 than a low-mass WD.

\subsection{Non-detection with the Fermi telescope}
A large number of MSPs have been discovered by radio observations of point sources
first identified by the Large Area Telescope (LAT) on board the \textit{Fermi} 
Gamma-Ray Space Telescope \citep[e.g.][]{ransom2011, keith2011b, cognard2011}.
We searched the Fermi LAT Second Source Catalog \citep{abd13} for point sources which 
could be associated with the five newly-discovered pulsars presented here, but found
no matches. Pulsar detections with \textit{Fermi} are usually parameterized in terms of
$\log(\sqrt{\dot{E}}/d^2)$ \citep[for a spin-down energy loss, $\dot{E}$, measured in 
erg~s$^{-1}$ and distance, $d$, in kpc; see\,][]{fermipsrcat}, which is usually $\gtrsim 17$
for pulsars that are detected. For PSR~J1431$-$4715, $\log(\sqrt{\dot{E}}/d^2) = 17$, 
so this pulsar might be considered to be right on the margins of detection by 
\textit{Fermi}, if indeed the DM distance is correct. The other MSPs presented here
all have values of $\log(\sqrt{\dot{E}}/d^2) \lesssim 16.5$.

\section{Possibility of optical detection for HTRU MSP companions} 
\label{s:opt}

Observations of companions of binary pulsars can allow the
determination of parameters which may not be measurable through pulsar
timing alone. Of particular interest here are millisecond pulsars with
low-mass helium-core white dwarf companions, where a dichotomy in the
thickness of the hydrogen envelope surrounding the helium-core of the
white dwarf leads to residual hydrogen burning, significantly slowing
down the cooling \citep{ashp96,dsbh98,asb01}. As a result, helium-core
white dwarfs with masses below approximately 0.2\,M$_\odot$ are typically
intrinsically brighter than higher mass white dwarfs. For those
systems where the white dwarf is bright enough for optical photometry,
it is possible to measure its temperature and cooling age, providing
independent constraints on the pulsar age
(e.g. \citealt{kbkk00,bkk03}). Furthermore, if the white dwarf has
suitable absorption lines, phase-resolved optical spectroscopy of the
white dwarf can be used to determine the mass-ratio and model the
white dwarf atmosphere to constrain both the white dwarf and pulsar
mass (e.g.\,\citealt{kbk96,cgk98,bkkv06,antoniadis2013}, and 
see \citealt{kbjj05} for a review).

In Table 4 we present predictions for the apparent brightness of helium-core
white dwarf companions to MSPs discovered in the HTRU survey. The
binary systems and their properties are listed in Table\,4. We use
predictions from white dwarf cooling models from \citet{bwb95}
and \citet{serenelli} to estimate the absolute magnitude by comparing the
predicted white dwarf cooling age with the characteristic spindown age
of the pulsar. The masses of the white dwarfs are constrained through
the timing measurements of the orbital period and projected semi-major
axis. Assuming a pulsar mass of 1.4\,M$_\odot$ and a range of probable
orbital inclinations, we obtain the masses listed in Table\,4.

Combining the observed pulsar dispersion measures with the NE2001
model of the Galactic electron distribution \citep{ne2001} yields the
distance estimates listed in Table\,4. Since some of the HTRU MSPs are
at low Galactic latitude, absorption can be significant. The Galactic
extinction model of \citet{al05} was used to obtain estimates of the
$V$-band absorption $A_V$ for the distance to and line-of-sight of
each MSP. Together, these provide the $V$-band distance modulus with
which the apparent magnitude can be estimated.

Figure\,6 combines all these estimates; the characteristic spindown
age $\tau_\mathrm{c}$ of the pulsar with the predicted cooling age of
the white dwarf. As pulsar spindown ages may not be a reliable age
estimator for the pulsar (Tauris 2012) we conservatively compare a
pulsar spin down ages from $0.1\tau_\mathrm{c}$ to
$\tau_\mathrm{c}$. The white dwarf cooling models then predict the
absolute $V$-band magnitude $M_V$. Combining these with the $V$-band
distance modulus $(m-M)_V$ gives the predicted apparent $V$-band
magnitude $m_V$. These estimates are also listed in Table\,4. 

Optical detection in imaging observations with an 8\,m class telescope
typically requires an apparent magnitude less than 24. For white
dwarfs brighter than 23rd magnitude optical spectroscopy may be
feasible provided suitable absorption lines are present in the
spectrum. Based on the estimates from Table\,4 and Fig.\,5, the HTRU
MSPs expected to have the white dwarf companions that can be detected
with 8\,m class telescopes are PSRs\,J1056$-$7117, 
J1431$-$4715 and J2236$-$5527, especially if the characteristic pulsar
age is an over-estimate of the real age of the system. In the case of 
PSR~J1125$-$5825, a detection may be possible if the characteristic age
vastly over-estimates the system's age.

\begin{figure}
	\begin{center}
	\includegraphics[width=8.5cm]{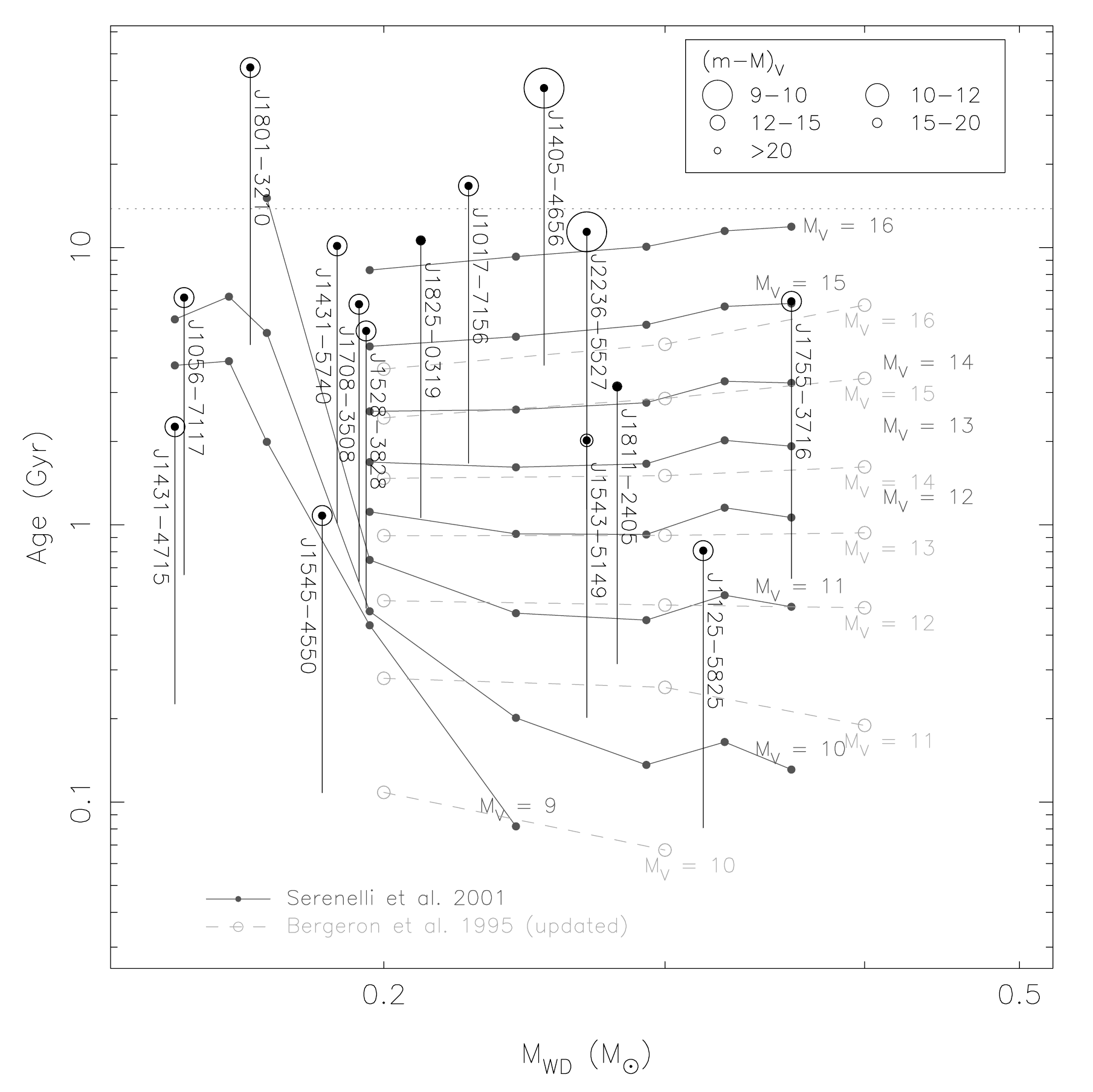}
	\end{center}
     \caption{Comparing characteristic pulsar spindown ages with
        white dwarf cooling ages. Predictions of white dwarf cooling age as a
        function of white dwarf mass following the models by \citet{bwb95}
        (dashed lines/open circles) and \citet{serenelli} (solid lines/closed
        circles) are shown for a set of absolute $V$-band magnitudes
        $M_V$. Plotted with the filled circles are the estimated companion
        masses $M_\mathrm{c,med}$ (assuming a 1.4\,M$_\odot$ pulsar and a
        $i=60^\circ$ inclination) against the characteristic spindown age
        $\tau_\mathrm{c}$ of the pulsar. The vertical line extends from the
        circles down to $0.1 \tau_\mathrm{c}$ to denote a conservative
        range of pulsar ages. Finally, the size of the circles concentric with
        the dots denoting the characteristic pulsar age and companion mass,
        scales with the $V$-band distance modulus $(m-M)_V$, where larger
        circles indicate lower distance moduli. From the distance modulus and
        the absolute $V$-band magnitude the apparent $V$-band magnitude for
        each companion for a given characteristic age can be calculated as
        $m_v=M_V+(m-M)_V$. The Hubble time is indicated by the dotted horizontal 
        line at $\sim 14$~Gyr.}
	\label{f:ceescooling}
\end{figure}

\section{Summary}
\label{s:discuss}

The discovery of five new MSPs from the High Time Resolution Universe survey has 
been presented.  These pulsars represent members of a wide range of known recycled
 pulsar types, including possibly the heaviest white dwarf known in orbit around a
 neutron star, 
 an eclipsing redback system, and an isolated, mildly recycled pulsar, 
indicating that it probably came from a HMXB which was disrupted during the second ccSN. 

Only one of these MSPs, PSR J1227$-$6208, is within the  
PMPS region.  This pulsar was in fact co-discovered
 in recent re-analyses of the PMPS data \citep{mlb+12,kek+13}. The superior 
temporal and spectral resolution of HTRU data means that even
with shorter pointings (in the 
mid-latitude region) than PMPS, HTRU is able to discover MSPs initially
 missed by previous surveys which used similar instrumentation
 to PMPS \citep{ebsb01,jbo+06,bjd+06}.  

It would be interesting to attempt detections of some WDs in the HTRU sample in order 
to test the reliability of MSP spin-down ages.  PSRs J1056$-$7117, J1125$-$5825, 
J1431$-$4715, and J2236$-$5527 are the most likely to result in positive detections.  
An optical observation of PSR J1431$-$4715 would also be informative 
for other reasons; if it is a redback 
system then its companion may be non-degenerate and it instead could be at the end
 of the spin-up phase \citep{rob11}.  The companion may also be distinctly 
non-spheroidal, and measurements of an orbitally modulated light curve can constrain
 the inclination \citep{svbk01,rcf+07}.  If spectroscopy as a function of orbital 
phase is possible for the companion then component masses can be constrained 
\citep{vbk11,rfs+12}.

\section*{ACKNOWLEDGEMENTS}
The Parkes Observatory is part of the Australia Telescope which is funded by the Commonwealth of Australia for operation as a National Facility managed by CSIRO. We thank the
reviewer, Dipankar Bhattacharya, for suggestions which helped improve the manuscript.

\bibliography{journals,allrefs,psrrefs,psrrefsBS,modrefsthornton}
\bibliographystyle{mnras}

\label{lastpage}
\end{document}